\documentclass[aps, prd, twocolumn, superscriptaddress, nofootinbib]{revtex4}
\usepackage{graphicx}
\usepackage{dcolumn}
\usepackage{amssymb}
\usepackage{amsmath,amssymb,amsfonts}

\begin{document}

\newcommand{\Eq}[1]{\mbox{Eq. (\ref{eqn:#1})}}
\newcommand{\Fig}[1]{\mbox{Fig. \ref{fig:#1}}}
\newcommand{\Sec}[1]{\mbox{Sec. \ref{sec:#1}}}

\newcommand{\PHI}{\phi}
\newcommand{\PhiN}{\Phi^{\mathrm{N}}}
\newcommand{\vect}[1]{\mathbf{#1}}
\newcommand{\Del}{\nabla}
\newcommand{\unit}[1]{\;\mathrm{#1}}
\newcommand{\x}{\vect{x}}
\newcommand{\y}{\vect{y}}
\newcommand{\p}{\vect{p}}
\newcommand{\ScS}{\scriptstyle}
\newcommand{\ScScS}{\scriptscriptstyle}
\newcommand{\xplus}[1]{\vect{x}\!\ScScS{+}\!\ScS\vect{#1}}
\newcommand{\xminus}[1]{\vect{x}\!\ScScS{-}\!\ScS\vect{#1}}
\newcommand{\diff}{\mathrm{d}}

\newcommand{\be}{\begin{equation}}
\newcommand{\ee}{\end{equation}}
\newcommand{\bea}{\begin{eqnarray}}
\newcommand{\eea}{\end{eqnarray}}
\newcommand{\vu}{{\mathbf u}}
\newcommand{\ve}{{\mathbf e}}


\title{Reappraisal of a model for deformed special relativity}

\newcommand{\addressImperial}{Theoretical Physics, Blackett Laboratory, Imperial College, London, SW7 2BZ, United Kingdom}
\newcommand{\addressRoma}{Dipartimento di Fisica, Universit\`a ‚ÄúLa Sapienza‚Äù
and Sez. Roma1 INFN, P.le A. Moro 2, 00185 Roma, Italia}

\author{Giulia Gubitosi}
\affiliation{\addressImperial}
 \author{Jo\~{a}o Magueijo}
\affiliation{\addressImperial}

\date{\today}

\begin{abstract}
We revisit one of the earliest proposals for deformed dispersion relations in the light of recent results on dynamical dimensional reduction and production of cosmological fluctuations. Depending on the specification of the measure of integration and addition rule in momentum space the model may be completed so as to merely
deform Lorentz invariance, or so as to introduce a preferred frame.  
Models which violate Lorentz invariance 
have a negative UV asymptotic dimension and a very red spectrum of quantum vacuum fluctuations. Instead, 
models which preserve frame independence can exhibit running to a UV dimension of 2, and a scale-invariant 
spectrum of fluctuations. The bispectrum of the fluctuations is another point of divergence between the two 
casings proposed here for the original model. 
\end{abstract}


\maketitle


\section{Introduction}
Deformed special relativity (DSR) was proposed as a simple way to accommodate modified dispersion relations (MDRs) without introducing
preferred frames~ \cite{AmelinoCamelia:2000mn, AmelinoCamelia:2000ge, KowalskiGlikman:2001gp, Magueijo:2001cr, Magueijo:2002am, KowalskiGlikman:2002ft}. 
One of the first DSR models was that proposed in~\cite{Magueijo:2001cr}, inspired by its position space counterpart,
initially  introduced by Fock~\cite{fock,manida}.
Given recent developments in the field, particularly with regards to cosmological perturbation generation~\cite{DSRflucts1,DSRflucts2,DSRrainbow,DSRmeasure} and the phenomenon of running of dimensionality~\cite{Ambjorn:2005db,Lauscher:2005qz,Horava:2009if,Modesto:2008jz,Carlip:2009kf,Benedetti:2008gu,Alesci:2011cg,Benedetti:2009ge,Calcagni:2010pa,Magliaro:2009if,Alkofer:2014raa,Calcagni:2014cza,Arzano:2014sya,Arzano:2014jfa,V:2015yma}, it is timely to reappraise this model.  
Some of its original motivations are now obsolete; however, exciting developments are now driving the field. The model remains one of the simplest of its kind, and it is interesting to revisit it in the light of recent work. 

We recall that the model proposed in~\cite{Magueijo:2001cr} is based upon the MDR:
\be\label{mdr1}
\frac{E^2-p^2}{(1-\lambda E)^2}=m^2
\ee
with a number of variations considered in~\cite{Magueijo:2002am} (specifically in order to implement a varying speed of light~\cite{vsl,vsl1,vsl-review}). This MDR was supplemented by a set of non-linear transformation laws between inertial observers, chosen in order to preclude the introduction of a preferred frame right at ``step one''. However 
the integration measure in momentum space for the model was never specified.   Depending 
on this ``step 2'' issue, the model may either be an example of  Lorentz invariance violation (LIV) or of proper relativistic DSR. The authors clearly intended the latter, but the 
potential for the model to be included in both frameworks should not be neglected. 

As a ``step 3'' issue we note that the energy-momentum addition rule and the mutiparticle sector of the theory add yet another layer where one may choose LIV or not. The early proposal in~\cite{Magueijo:2002am} 
accommodates the latter, but a linear addition rule would also allow for a LIV element in the framework.  
In this paper we consider the various completions of this model in view of these possibilities.
The ``step 2''  options will be seen to have dramatic implications for the 
asymptotic UV dimensionality of the space-time and the spectrum of cosmological fluctuations. ``Step 3''
issues will be found to leave an imprint onto the bispectrum and higher-order correlators of the fluctuations.

\section{To LIV or not no LIV}\label{LILIV}
The MDR (\ref{mdr1}) does not by itself specify whether the theory has broken Lorentz invariance or not, even if
transformation laws between observers are postulated such that the MDR remains invariant. As explained in ~\cite{ Magueijo:2001cr} 
we should select a non-linear representation of the Lorentz group if we do not want to introduce a preferred frame right from the start
(as would be the case if linear transformations were chosen). This amounts to postulating standard transformations for
the auxiliary ``linearizing'' variables:
\bea
{\tilde E}&=&\frac{E}{1-\lambda E}\\
{\tilde p_i}&=&\frac{p_i}{1-\lambda E}
\eea
resulting in transformations such that a deformed boost along the $\hat z$ direction, with velocity $v$, reads 
\bea
E'&=&{\xi\left(E- vp_z \right) 
\over 1+\lambda (\xi -1) E  -\lambda\xi v p_z }\nonumber\\  
p_z'&=&{\xi\left(p_z- v E\right) 
\over 1+\lambda (\xi -1) E  -\lambda\xi v p_z }\nonumber\\  
p_x'&=&{p_x
\over 1+\lambda (\xi -1) E  -\lambda\xi v p_z }\nonumber\\  
p_y'&=&{p_y
\over 1+\lambda (\xi -1) E  -\lambda\xi v p_z }\label{transf}
\eea
where $\xi=1/\sqrt{1-v^2}$ (assuming $D=3$ spatial dimensions). These are the transformation rules for the physical variables.  It is assumed that the interactions peg down the physical frame as the MDR frame, because only in this frame is there minimal coupling (or Einstein gravity is valid).

But even if care has been taken to prevent the introduction of a preferred frame at line one, further work is required to prevent it from sneaking in elsewhere.  For instance, one must specify the measure of integration in momentum space, a matter closely related to the identification of the dimensionality of spacetime. 
We clearly introduce a preferred frame if we supplement (\ref{mdr1}) and (\ref{transf})
with an undeformed measure in momentum space:
\be\label{muLIV}
d\mu=dE\, dp\, p^{D-1}.
\ee
In contrast we do not introduce a preferred frame if we choose measure: 
\be\label{muLI}
d\mu=\frac{dE\, dp\, p^{D-1}}{(1-\lambda E)^{D+2}},
\ee
since this is invariant under transformations (\ref{transf}), as can be inferred from the fact that in 
linearizing coordinates this measure becomes:
\be\label{DSRtildemu}
d\tilde\mu=d{\tilde E}\, d{\tilde p}\, {\tilde p}^{D-1},
\ee
i.e. the trivial measure, which is invariant under linear transformations. 

The composition rule for energy and spatial momenta is another important element where one may have LIV or not. 
If a linear composition rule is specified for $E$ and $p$, then this obviously selects a preferred frame. However, if the integration measure (\ref{muLI}) has been chosen, the preferred frame is evident only at the level of the multiparticle sector (and interactions), with the preferred frame silent when examining free particles. This shows the multitude of options available, as we could also combine a linear addition rule with measure (\ref{muLIV}), resulting in a fully LIV theory (i.e. one where the preferred frame is manifest at all levels).

To avoid the introduction of a preferred frame at all a number of possibilities were considered in~\cite{Magueijo:2001cr}. The deformed sum $\oplus$ of the momenta of $N$ particles may be chosen to be:
\bea
E_{1}\oplus ....\oplus E_{N}&=&\frac { \sum_{n}\frac{ E_{n}}{1-\lambda E_{n}}}{1+f_N(\lambda) \sum_{n}\frac{E_{n}}{1-\lambda E_{n}}} \label{eq:Esum}\\
\vec p_{1}\oplus ....\oplus \vec p_{N}&=&\frac { \sum_{n}\frac{ \vec p_{n}}{1-\lambda  E_{n}}}{1+f_N(\lambda) \sum_{n}\frac{E_{n}}{1-\lambda E_{n}}} .\label{eq:psum}
\eea
The function $f_N(\lambda)$ may be non-trivial if we wish the multiparticle energy-momentum to transform differently from the single particle one. In fact, the multiparticle transformations then mimic the single particle ones, but with $\lambda$ replaced by $f_N(\lambda)$. This was proposed to circumvent the ``soccer ball'' problem, and at its simplest $f_N=\lambda/N$. It leads to a commutative but non-associative addition rule; in fact, we must have: 
\be
p_1\oplus p_2\oplus p_3\neq (p_1\oplus p_2)\oplus p_3\neq p_1\oplus (p_2\oplus p_3).
\ee
However, the soccer ball is by and large a red herring~\cite{Amelino-Camelia:2013fxa, Amelino-Camelia:2014gga}, and it never exists in field theory~\cite{lasers,sabine,sabine1}. So we can simply set $f_N=\lambda$, leading to a commutative and associative addition law, without introducing a preferred frame. All of these choices regarding addition rules will have signatures in the bispectrum of the theory.

\section{Spacetime dimensionality and vacuum fluctuations}\label{dimPK}
The implications for the dimensionality of space-time of the various LIV/DSR options for completing the model are dramatic. 
The phenomenon of running of the dimensionality of spacetime, as one scans from IR to UV, has been widely studied~\cite{Ambjorn:2005db,Lauscher:2005qz,Horava:2009if,Modesto:2008jz,Carlip:2009kf,Benedetti:2008gu,Alesci:2011cg,Benedetti:2009ge,Calcagni:2010pa,Magliaro:2009if,Alkofer:2014raa,Calcagni:2014cza,Arzano:2014sya,Arzano:2014jfa,V:2015yma}. Its most direct relation to MDRs was exposed in~\cite{DSRmeasure,dimredLI,ADSDSR}, where it was shown that the phenomenon may be characterized in terms of the Haussdorf dimension of energy-momentum space in units such that all non-trivial effects are shifted to the integration measure. This always agrees {\it asymptotically} with results obtained using the spectral dimension as a tool, but bypasses many of its conceptual drawbacks~\cite{Calcagni:2013vsa}. 

For our model, in the LIV case, we can write the measure (\ref{muLIV}) in 
terms of linearizing variables resulting in:
\be
d\tilde \mu=\frac{d\tilde E\, \tilde p^{D-1}\, d\tilde p}{(1+\lambda\tilde E)^{D+2}}
\rightarrow \frac{d\tilde E\, \tilde p^{D-1}\, d\tilde p}{(\lambda\tilde E)^{D+2}}
\ee
where the last limit represents the UV regime. Energy and spatial momentum have UV dimensionality:
\bea
D_E&=&-D-1\\
D_p&=&D
\eea
and therefore total Haussdorf dimension $d_H=-1$, independently of $D$. It is not clear that a negative dimensionality of energy-momentum space is pathological. It merely signals that, as we consider momentum space shells located further and further outwards, we find that the number of modes they contain falls off, instead of increasing. The number of modes contained in these shells is {\it not} negative (something which would obviously  be pathological). A negative dimensionality is therefore a theoretical possibility. 

Regardless of this issue, it is obvious that the model is phenomenologically problematic. As explained in~\cite{VacuumFluctuations} the spectrum of quantum vacuum fluctuations generated in the UV is directly related to the dimension of momentum space. Our result implies that in this model these fluctuations have a very red spectrum, with: 
\be
n_S=-2.
\ee
With basic assumptions on gravity, this is also the spectrum left outside the horizon, a conclusion which is independent of $D$. Of course some assumptions may be modified, but in the light of previous work~\cite{DSRflucts1,DSRflucts2,DSRrainbow,DSRmeasure,VacuumFluctuations,GRFessay}
our result suggests that the LIV completion of the model is phenomenologically pathological. We stress that by replacing the Casimir of the theory by a power of itself (as in ${\cal C}\rightarrow {\cal C}^{1+\gamma}$) these conclusions would not change. The UV dimension would still  be negative (indeed, $d_H=-1/(1+\gamma)$), and the spectrum of fluctuations be red.

If we choose a DSR measure, however, the situation is very different. Since in linearizing units (where all non-trivial effects are shifted to the measure) we have  (\ref{DSRtildemu}), there is no running of the dimensionality in this model. However, just as for de Sitter momentum space~\cite{dimredLI}, anti-de Sitter momentum space~\cite{ADSDSR}, and what in~\cite{VacuumFluctuations} was called non-local Lorentz invariant theories, we can consider
the class of associated models with MDRs
\be
\frac{E^2-p^2}{(1-\lambda E)^2}+\lambda^{2\gamma} \left[\frac{E^2-p^2}{(1-\lambda E)^2}\right]^{1+\gamma}=m^{2}\,.\label{eq:gammaMDR}
\ee
In the UV limit this becomes 
\be\label{mdr2}
\lambda^{2\gamma}\left[\frac{E^2-p^2}{(1-\lambda E)^2}\right]^{1+\gamma}\approx m^{2}.
\ee
Just as in that case we then find that the UV spectral dimension is:
\be
d_H=\frac{1+D}{1+\gamma}
\ee
so that for $D=3$ and $\gamma=1$ we run to 2 dimensions in the UV, and so~\cite{VacuumFluctuations}  the vacuum fluctuations have a power spectrum with the coveted:
\be 
n_S=1,
\ee
with interesting phenomenological implications~\cite{GRFessay}.
A subtlety similar to that addressed in~\cite{kPfieldTheory} must be resolved here, too, and will be addressed in the next Section.


\section{The sum rule and the bispectrum}
Given the various composition rules proposed in Section~\ref{LILIV}, a number of possibilities emerge regarding the density fluctuations. This is relevant both for the details of the evaluation of the power spectrum, as well as the selection rules constraining the higher order correlators. 

If the sum rule is linear (and thus we have a LIV theory), then, first of all, the argument in~\cite{VacuumFluctuations} for evaluating the power spectrum does apply directly, without further refinement. In addition, the selection rules for the bispectrum (and higher order correlators) are trivial. For example the bispectrum must take the form:
\be
\langle \phi^{3}\rangle \propto \delta^{(3)}\left(\vec k_{1} + \vec k_{2} + \vec k_{3}\right)F(\vec k_{1},\vec k_{2},\vec k_{3})\,,
\ee
i.e. be non-vanishing only on triangles. The function $F(\vec k_{1},\vec k_{2},\vec k_{3})$ depends on the size and shape of the triangles and is determined by the details of the interaction Hamiltonian of the theory. 

The situation is different should we use a DSR sum rule, such as (\ref{eq:Esum}) and (\ref{eq:psum}). Then, care must be taken to make sure one does not introduce a preferred frame at quantization, by applying the prescription proposed in~\cite{VacuumFluctuations}, as explained in~\cite{kPfieldTheory}. Proper, frame independent quantization follows from identifying the ``antipode'' in momentum space. For this 
to be compatible with Eqs. (\ref{eq:Esum}) and (\ref{eq:psum}) (with $f_N=\lambda$), the antipode, defined as
\bea
E\oplus (\ominus E)&=&0 \nonumber \\
\vec p\oplus (\ominus \vec p)&=&0\nonumber\\
\eea
should be given by:
\bea
\ominus E&=&-\frac{E}{1-2 \lambda E}\\
\ominus \vec p&=&-\frac{\vec p}{1-2 \lambda E}\,.
\eea
The two-point function is given by the single-particle state normalization, induced by $\delta(p\oplus (\ominus p')$. If $\gamma=0$, in linearizing coordinates both the measure and the dispersion relation are the trivial ones, as well as the conservation laws of momenta. So the two-point function will be the standard one, and no preferred frame has been introduced at quantization, as long as we restrain ourselves to the 2-point function.  However, if $\gamma\neq 0$, a further linearization procedure is needed, equal in everything to that found for non-local Lorentz invariant theories in~\cite{VacuumFluctuations} (see  Section VA). Specifically, for $\gamma=1$ and $D=3$ we should linearize via a further step, first by introducing coordinates:
\bea
\tilde E&=&\tilde r\cosh \tilde\phi\\
\tilde p&=&\tilde r\sinh \tilde\phi
\eea
then 
\bea
\hat r&=&\tilde r^2\\
\hat \phi&=&\frac{1}{6}\tilde \phi
\eea
and finally: 
\bea
\tilde E&=&\tilde r\cosh \tilde\phi\\
\tilde p&=&\tilde r\sinh \tilde\phi.
\eea
The measure in linearizing units will therefore  be
\be
d\hat \mu=d\hat E \, d\hat p
\ee
and this does correspond to the conclusion that $n_S=1$, as explained in~\cite{VacuumFluctuations}, and likewise it does not introduce a preferred frame.

The results for the two-point function are therefore trivial unless $\gamma\neq0$, for the DSR theory. 
This is because in the absence of interactions, and as long as we do not look at the spacetime counterpart of the theory, we cannot distinguish between different momentum space frames connected by non-linear redefinitions of momenta. When $\gamma=0$  our DSR model is connected to the standard special-relativistic theory by a non-linear redefinition of momenta. 
Once interactions are introduced, however, the momentum frame is fixed. This explains why the three-point function has  non-trivial features, even when $\gamma=0$, as we now show.

In general the three-point function is of the form:
\be
\langle \phi^{3}\rangle \propto \delta^{(3)}\left(\vec k_{1}\oplus \vec k_{2}\oplus \vec k_{3}\right)F(\vec k_{1},\vec k_{2},\vec k_{3})\,,
\ee
where the amplitude $F$ depends on the details of the interactions, and the momenta $\vec k_{1}, \vec k_{2}, \vec k_{3}$ are on-shell.
If the theory is written in linearizing coordinates, then the delta function contains the standard sum of momenta and it constrains these momenta to form triangular shapes.
If, however, we consider the theory in the MDR frame, the delta function of the deformed sum of momenta does not in general constrain the momenta to form triangular shapes. This signals the non-equivalence of the two frames, as selected by the interaction Hamiltonian. 

If we take (\ref{eq:Esum}) and (\ref{eq:psum}) with $f_N(\lambda)=\lambda$, then
in the MDR frame the argument of the delta function is: 
\bea
\vec k_{1}\oplus \vec k_{2}\oplus \vec k_{3} &=&
\frac{\frac{\vec k_{1}}{1-\lambda |\vec k_{1}|}+\frac{\vec k_{2}}{1-\lambda |\vec k_{2}|}+\frac{\vec k_{3}}{1-\lambda |\vec k_{3}|}}{1+\lambda \left(\frac{\vec k_{1}}{1-\lambda |\vec k_{1}|}+\frac{\vec k_{2}}{1-\lambda |\vec k_{2}|}+\frac{\vec k_{3}}{1-\lambda |\vec k_{3}|}\right)}\nonumber
\eea
(were we have taken into account the massless on-shell relation $E_{i}=|\vec k_{i}|$ ). 
It is clear that the condition $\vec k_{1}\oplus \vec k_{2}\oplus \vec k_{3}=0$ is equivalent to  $\vec k_{1}+ \vec k_{2}+ \vec k_{3}=0$ in only two cases: 
\begin{itemize}
\item 
if one of the vectors is zero and the other two are of equal length but opposite direction (i.e. we have a squeezed triangle), 
\item
if $|\vec k_{1}|=|\vec k_{2}|=|\vec k_{3}|$ (i.e. we have an equilateral triangle). 
\end{itemize}
In all the other cases the condition $\vec k_{1}\oplus \vec k_{2}\oplus \vec k_{3}=0$ leads to an ``open triangle''. To see this we should write the deformed sum as a standard sum plus a correction term:
\bea
\vec k_{1}\oplus \vec k_{2}\oplus \vec k_{3}&=&\vec k_{1}+ \vec k_{2}+ \vec k_{3} \nonumber\\
&+&\lambda \left(\frac{|\vec k_{1}| \vec k_{1}}{1-\lambda |\vec k_{1}|}+\frac{|\vec k_{2}|\vec k_{2}}{1-\lambda |\vec k_{2}|}+\frac{|\vec k_{3}|\vec k_{3}}{1-\lambda |\vec k_{3}|}\right).\nonumber
\eea
It is then easy to check that if we require both $\vec k_{1}+ \vec k_{2}+ \vec k_{3}$ and the term in parentheses to be zero one of the two aforementioned conditions has to hold. Otherwise, the term in parentheses fails to vanish and gives the size of the ``gap'' left for the triangle to close.

The selection rule just discovered is to be contrasted with that found for the LIV completion of the model (closed triangles only), and that identified in~\cite{kPfieldTheory} for de Sitter momentum space (squeezed triangles only; a different spectrum of gaps for the open triangles). This  highlights the discriminating potential of the bispectrum.

\section{Conclusions}
In this paper we reexamined one of the earliest models for MDRs in the light of recent developments in the field. We found the original formulation of the model to lack sufficient information in order to address the relevant issues, specifically regarding the integration measure in momentum space and the addition rule of energy and momentum. We proposed LIV and DSR measures of integration and addition rules, leading to at least 4 combinations of possibilities, all with very different theoretical and phenomenological implications. 

We found that a LIV integration measure leads to running to negative dimension in the UV and a strongly red spectrum of density fluctuations. In contrast, a DSR integration measure may run to 2 UV dimensions, specifically if the most obvious Casimir is squared. This is associated with the production of scale-invariant vacuum quantum fluctuations. 

In addition, it is possible to supplement a DSR measure with a LIV addition rule, resulting in Lorentz violation only at the level of the multiparticle sector and the 3-point function and higher order correlators. For the theory to be fully relativistic a non-trivial addition rule must be specified. This results in a non-vanishing bispectrum for open triangles with very specific gaps depending on size and angle. The only  exceptions are equilateral and squeezes triangles, on which the bispectrum is still non-vanishing. This is an interesting signature of these theories. 

In conclusion, 
if we accept a DSR completion of the original model, then it  differs very little from the non-local Lorentz invariant theory considered in~\cite{VacuumFluctuations} and elsewhere in the literature. In that model the standard wave operator is squared. In both cases there is running to 2 dimensions in the UV and the vacuum fluctuations are scale-invariant. This is not surprising, since in both cases  momentum space is actually flat. The theories are different, however, if we examine interactions and the bispectrum. Then the different addition rules selected by the different interaction Hamiltonians will bring to the fore fundamental differences, as we have shown.

The DSR model considered here, however, is always very different from the de Sitter momentum space model. This boils down to the intrinsic curvature of the momentum space of that model, but also its group theory structure (for example, the Casimir operator is different from the metric invariant). All of these theories provide a rich Planck scale phenomenology regarding running of dimensionality and the cosmological density fluctuations. They should be considered on an equal footing until an experimental clue comes to the rescue. In this paper we pointed to a possible location where to look.

\section{Acknowledgments}
We  acknowledge support from the John Templeton Foundation. JM was also funded by a STFC consolidated grant and the Leverhulme Trust.

\end{document}